\documentclass[conference]{IEEEtran}
\IEEEoverridecommandlockouts

\usepackage{amsmath, amssymb}             
\usepackage{amsthm}

\newtheorem{theorem}{Theorem}
\newtheorem{lemma}{Lemma}
\newtheorem{proposition}{Proposition}
\renewcommand{\IEEEproofindentspace}{0pt}  

\usepackage[
  range-phrase=--,
  per-mode=symbol-or-fraction,
  range-units=single,
  list-units=single,
  detect-all,
  list-final-separator={, and },
]{siunitx}

\usepackage{graphicx}

\usepackage{xcolor}

\usepackage{tikz}
\usetikzlibrary{decorations.pathreplacing}

\usepackage{booktabs}

\usepackage{cite}
\usepackage[hidelinks,bookmarks=false]{hyperref}
\hypersetup{pdftitle={Survival-Time Risk Beyond Low-Order Channel Models: Identifiability and Scheduling}, pdfauthor={Ruifeng Zheng, Zhihan Xu, Pit Hofmann, and Frank H. P. Fitzek}}
\usepackage[capitalise,noabbrev]{cleveref}
\Crefformat{figure}{#2Fig.~#1#3}
\Crefformat{equation}{#2Eq.~(#1)#3}
\crefname{figure}{Fig.}{Figs.}
\crefname{equation}{Eq.}{Eqs.}
\crefname{section}{Sec.}{Secs.}
\Crefname{section}{Sec.}{Secs.}

\begin{document}
\bstctlcite{IEEEexample:BSTcontrol}
\title{Survival-Time Risk Beyond Low-Order Channel Models: Identifiability and Scheduling}

\author{
    \IEEEauthorblockN{
        Ruifeng Zheng\IEEEauthorrefmark{1},
        Zhihan Xu\IEEEauthorrefmark{1}, 
        Pit Hofmann\IEEEauthorrefmark{1}\IEEEauthorrefmark{2}, and
        Frank H.\,P. Fitzek\IEEEauthorrefmark{1}\IEEEauthorrefmark{2}
    }
    \IEEEauthorblockA{
         \IEEEauthorrefmark{1}Deutsche Telekom Chair of Communication Networks, Dresden University of Technology, Germany\\
         \IEEEauthorrefmark{2}Centre for Tactile Internet with Human-in-the-Loop (CeTI), Dresden, Germany\\
         \texttt{\{ruifeng.zheng, zhihan.xu, pit.hofmann, frank.fitzek\}@tu-dresden.de}
    }
    \vspace{-0.3cm}
    \thanks{This work was funded by the German Research Foundation (DFG, Deutsche Forschungsgemeinschaft) as part of Germany’s Excellence Strategy – EXC 2050/2 – Project ID 390696704 – Cluster of Excellence “Centre for Tactile Internet with Human-in-the-Loop” (CeTI) of Dresden University of Technology.
    The authors also acknowledge the financial support by the Federal Ministry of Research, Technology and Space (BMFTR) of Germany in the program “Souverän. Digital. Vernetzt.” Joint project 6G-life, project identification number 16KIS2413K, the program “Verbundprojekt: Disruptive Kommunikationsparadigmen für technologische Souveränität, Resilienz und Shared Prosperity - Translation in Industrie und Aufbau innovativer Technologiedemonstratoren - CommUnity,” project identification number 16KISS012K.}
}

\maketitle

\begin{abstract}
This paper investigates whether low-order packet-loss statistics suffice to assess survival-time risk under bursty packet losses, i.e., the probability that no update succeeds for a prescribed number of consecutive slots. Conventional models, such as the Gilbert model, characterize packet losses using the loss rate and mean burst length, but channels sharing these statistics may exhibit substantially different failure risks. We characterize the Markov channels one order above a given low-order fit that share its statistics and analyze their risks under acknowledgment-based scheduling with transmission budgets. We show that, with dense transmission attempts, risk is primarily governed by loss persistence beyond the modeled memory, and the worst-case channel need not be the most persistent one. Under limited budgets, retry spacing makes loss relapse probabilities important and can reverse the risk ordering of compatible channels. Numerical and measurement results show that low-order models can substantially underestimate the risk, highlighting the need for scheduling-dependent channel statistics.
\end{abstract}

\begin{IEEEkeywords}
Survival time, bursty packet loss, Markov channel models, risk identifiability, transmission scheduling, model uncertainty.
\end{IEEEkeywords}

\section{Introduction}
\label{sec:intro}

\IEEEPARstart{I}{ndustrial} control and human-in-the-loop applications of the Tactile Internet, such as teleoperation, require timely and reliable information updates~\cite{fitzek2021tactile}. A prolonged interruption in successful packet delivery may violate the application's survival-time requirement~\cite{3gpp22104}, even when the average packet-loss rate is low. Assessing reliability therefore requires accounting for temporal correlations in packet losses rather than their average rate alone. In practice, bursty communication channels are commonly represented by low-order models, such as the Gilbert model~\cite{gilbert1960} and first-order Markov models of fading~\cite{zorzi1998}, which provide tractable descriptions of packet losses using a small number of statistics, including the loss rate and mean burst length. However, different channels may share these statistics while exhibiting substantially different probabilities of survival-time violations.

The temporal structure of packet losses and its effect on reliability have been widely investigated~\cite{hossler2018,scheuvens2021,singh2026}. Previous studies have examined Markov models of different orders using measured loss traces~\cite{yajnik1999} and queueing analysis~\cite{yu2008accuracy}, as well as scheduling and remote estimation over Markovian channels with acknowledgment feedback~\cite{yao2022aoi,dutta2023,wang2026pomab}. Related work has also characterized information freshness, age violations, and estimation stability under correlated losses~\cite{wu2018ge,guan2022memory}. Robust verification and control methods~\cite{quatmann2016,wiesemann2013} and reliability bounds under unknown dependence~\cite{ge2026} can account for channel-model uncertainty. However, these approaches assume a given model or uncertainty set, not the set of channels a low-order fit cannot resolve. A fundamental question remains: \emph{to what extent can survival-time risk be determined from low-order channel statistics, and which additional statistics are needed for reliable risk assessment?}

This paper addresses this question by considering channels that share the same low-order packet-loss statistics but differ in their higher-order dependencies. We investigate their survival-time risks under acknowledgment-based scheduling with a finite transmission budget. The key observation is that the relevant channel statistics depend on the transmission schedule. When transmissions are attempted in every slot, consecutive packet losses primarily determine the failure risk. Under a limited transmission budget, however, successful channel opportunities during idle slots cannot reset the age of information. Consequently, the spacing between transmission attempts introduces additional dependence on loss relapses, potentially changing the relative reliability of channels with identical low-order statistics.

The main contributions of this paper are as follows.
(i) We characterize the Markov channels one order above a low-order fit as an explicit box whose maximum-entropy point is the fitted model.
(ii) We bound the risk of every channel in this box and show that, with dense attempts, loss persistence beyond the modeled memory governs the risk to first order, while the worst case need not be the most persistent channel.
(iii) We show that under a budget, retry spacing makes loss relapses matter, and construct two compatible channels whose risk order the budget reverses.
(iv) We show by exact computation, fading simulations, and measured IEEE 802.15.4 links that low-order models can substantially underestimate the risk, even with a second-order fit. Full proofs, code, and data are available in~\cite{zheng2026repo}. 

\section{System Model and Problem Statement}
\label{sec:model}

\subsection{Channel and Compatible Class}
\label{subsec:channel}

We consider a slotted communication link and denote by $\gamma_t\in\{0,1\}$ the potential packet-delivery outcome in slot $t$, where $\gamma_t=1$ indicates success and $\gamma_t=0$ indicates loss if a transmission is attempted. The latent channel evolves in every slot, including idle ones. We model $\{\gamma_t\}$ as a stationary Markov process of order at most $M$, with context $S_t=(\gamma_{t-M},\ldots,\gamma_{t-1})$ and conditional success probability
\begin{equation}
p(w)=\Pr(\gamma_t=1\mid S_t=w),
\qquad \eta\le p(w)\le 1-\eta,
\end{equation}
where $w\in\{0,1\}^M$ ranges over the contexts and $\eta\in(0,\frac12)$ is a known bound that excludes deterministic transitions. Equivalently, the channel is characterized by its stationary $(M+1)$-block distribution $\nu$, with $p(w)=\nu(w1)/\mu(w)$, where $wa$ denotes the word $w$ followed by $a\in\{0,1\}$ and $\mu(w)=\nu(w0)+\nu(w1)$ is the probability of context $w$. More generally, $\Pr(x)$ denotes the probability that consecutive outcomes form the word $x$; e.g., $\Pr(0^j)$ is the probability of $j$ consecutive losses.

Suppose a Markov model of order $L<M$ is fitted using the stationary $(L+1)$-block distribution $c$, i.e., $c(x)=\Pr(x)$ for words $x$ of length at most $L+1$. We assume $c$ is known so as to separate model ambiguity from estimation error. For $L=1$, these statistics determine the packet-loss rate and mean burst length, as in the Gilbert model. We define the \emph{compatible class} $\mathcal{C}_L(c;\eta)$ as the set of all stationary, $\eta$-positive Markov channels of order at most $M$ sharing the same $(L+1)$-block distribution $c$. These channels are indistinguishable under the low-order statistics but may differ in their higher-order dependencies. We focus on $M=L+1$, the smallest order at which such ambiguity arises, enabling an explicit characterization of the compatible class.

\subsection{Survival Time, Budget, and Risk}
\label{subsec:risk}

We consider an application that tolerates at most $D-1$ consecutive slots without a successful update, where $D$ denotes the \emph{survival-time threshold}~\cite{3gpp22104}. In remote estimation, for example, the estimation error covariance is $\bar P$ immediately after a successful update and evolves to $\Psi^d(\bar P)$ after $d$ slots without one, where $\Psi$ is the open-loop Lyapunov map~\cite{sinopoli2004}. If the application requires $\operatorname{tr}P\le P_{\max}$ and $\operatorname{tr}\Psi^d(\bar P)$ is nondecreasing in $d$, the survival-time threshold is
\begin{equation}
D=\min\{d:\operatorname{tr}\Psi^d(\bar P)>P_{\max}\},
\end{equation}
where $\operatorname{tr}(\cdot)$ denotes the trace and $P$ the estimation error covariance.

\begin{figure}[t]
\centering
\begin{tikzpicture}[font=\scriptsize,x=0.74cm,y=0.38cm]
\node[anchor=east] at (0.45,0) {latent $\gamma_t$};
\node[anchor=east] at (0.45,-1.25) {dense: $d_t$};
\node[anchor=east] at (0.45,-2.5) {every 2nd slot: $d_t$};
\foreach \i/\g in {1/1,2/0,3/1,4/0,5/0,6/1,7/0,8/1}{
  \ifnum\g=0 \filldraw[fill=gray!45,draw=black] (\i-0.38,-0.42) rectangle (\i+0.38,0.42);
  \else \draw (\i-0.38,-0.42) rectangle (\i+0.38,0.42); \fi
  \node at (\i,0) {\g};
  \node[gray] at (\i,-3.55) {\i};}
\node[gray,anchor=east] at (0.45,-3.55) {slot $t$};
\foreach \i/\d in {1/0,2/1,3/0,4/1,5/2,6/0,7/1,8/0}{\node[draw,circle,inner sep=0.8pt] at (\i,-1.25) {\d};}
\foreach \i/\d in {1/1,3/3}{\node at (\i,-2.5) {\d};}
\foreach \i/\d in {2/2,4/4}{\node[draw,circle,inner sep=0.8pt] at (\i,-2.5) {\d};}
\node[draw,very thick,inner sep=1.2pt] at (5,-2.5) {\textbf{5}};
\node[anchor=west] at (5.3,-2.5) {$=D$: \textbf{failure}};
\draw[decorate,decoration={brace,amplitude=3pt}] (1.62,0.55) -- (4.38,0.55) node[midway,above=2pt] {relapse $0\,1\,0$};
\end{tikzpicture}
\caption{Latent channel outcomes $\gamma_t$ and the resulting age $d_t$ under dense attempts and under attempts every second slot ($D=5$; shaded: loss; circled: attempted slot).}
\label{fig:system}
\vspace{-0.5cm}
\end{figure}
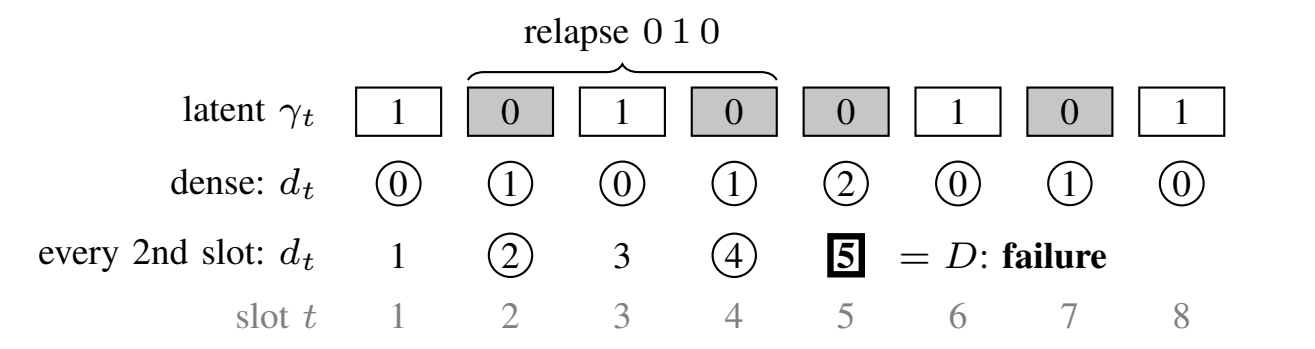

In each slot $t$, the scheduler at the transmitter selects $u_t\in\{0,1\}$, where $u_t=1$ indicates a transmission attempt. The scheduler has a budget of at most $B$ attempts (e.g., to save energy~\cite{yao2022aoi}) over a mission of $T$ slots, with the constraint enforced on every sample path. Let $d_t$ denote the number of slots since the last successful update, initialized as $d_0=0$. Its evolution is
\begin{equation}
d_t=
\begin{cases}
0, & u_t\gamma_t=1,\\
d_{t-1}+1, & \text{otherwise}.
\end{cases}
\end{equation}
For each attempt, the scheduler receives immediate and error-free acknowledgment (ACK/NACK) feedback, so it observes $\gamma_t$ only in slots with $u_t=1$. A success in an idle slot therefore goes unobserved and does not reset the age. In \cref{fig:system}, for example, the success in a loss--success--loss pattern resets the age under dense attempts, whereas under attempts every second slot it falls into an idle slot and the age reaches $D$.

The channel starts in its stationary regime, i.e., $S_1$ is drawn from $\mu$. A survival-time violation occurs when the age reaches $D$, which first happens at
\begin{equation}
\tau=\min\{t:d_t\ge D\}.
\end{equation}
The mission fails if this happens within its $T$ slots, so the mission failure risk of a scheduling policy $\pi$ under channel $\mathcal M$ is
\begin{equation}
R_{\mathcal M}(\pi)
=\Pr\nolimits_{\mathcal M}^{\pi}(\tau\le T),
\label{eq:risk}
\end{equation}
and $1-R_{\mathcal M}(\pi)$ is the corresponding mission reliability.

The scheduler does not know the true channel, only the fitted model. We denote by $\pi_L$ the policy that minimizes the mission failure risk under the fitted $L$-th order model, subject to the transmission budget. For $L=1$, the fitted model is characterized by the success probabilities after a loss and after a success and by their one-slot propagation,
\begin{equation}
q_o=\frac{c(o1)}{c(o)},\qquad
f(x)=q_0+(q_1-q_0)x.
\end{equation}
Since the scheduler observes the channel only when it transmits, it predicts a success probability of $f^{\delta-1}(q_o)$ from its most recent observation $o$, made $\delta$ slots earlier, where $f^j$ denotes the $j$-fold composition of $f$; before the first observation, it predicts $c(1)$. The policy $\pi_L$ then follows by finite-horizon dynamic programming over the state $(t,d,r,o,\delta)$, where $r$ is the remaining budget, with ties broken in favor of idling.

\subsection{Risk Identifiability}
\label{subsec:identifiability}

The scheduler designs $\pi_L$ from the low-order statistics $c$, but the true channel may be any member of $\mathcal{C}_L(c;\eta)$. The low-order statistics therefore determine the risk of $\pi_L$ only up to the interval $[R^-(c),R^+(c)]$, with
\begin{equation}
R^{\pm}(c)=\operatorname*{max/min}_{\mathcal M\in\mathcal{C}_L(c;\eta)}R_{\mathcal M}(\pi_L)
\end{equation}
the largest and smallest risks over compatible channels, whereas the fitted model predicts the nominal risk
\begin{equation}
R_{\mathrm{nom}}=R_{\mathrm{fit}}(\pi_L).
\end{equation}

This ambiguity becomes critical when the interval contains the risk threshold. For a prescribed threshold $\varepsilon$, if
\begin{equation}
R^-(c)\le\varepsilon<R^+(c),
\end{equation}
the same low-order statistics are consistent both with channels that meet the requirement $R_{\mathcal M}(\pi_L)\le\varepsilon$ and with channels that violate it, so whether the requirement is met is not identifiable from $c$. Our objective is to characterize this ambiguity and to identify the additional channel statistics that resolve it under different transmission schedules.

\subsection{Structure of the Compatible Class}
\label{subsec:box}

The risk ambiguity arises because a survival-time violation typically depends on the channel over a window of $D+1$ slots, more than the $L+1$ slots that $c$ describes. Over such windows, compatible channels can differ in higher-order dependencies that $c$ does not determine. For $M=L+1$, these dependencies are captured by the probabilities $x_m=\nu(0m0)$, $m\in\{0,1\}^L$, that two losses are separated by the pattern $m$.

\begin{lemma}[Box structure]\label{lem:box}
For $M=L+1$, the compatible class $\mathcal{C}_L(c;\eta)$ is affinely equivalent to $\prod_m[x_m^{\mathrm{lo}},x_m^{\mathrm{hi}}]$ in the coordinates $x_m$, where
\begin{align*}
x_m^{\mathrm{lo}}&=\max\{\eta c(0m),\,c(m0)-(1-\eta)c(1m)\},\\
x_m^{\mathrm{hi}}&=\min\{(1-\eta)c(0m),\,c(m0)-\eta c(1m)\}.
\end{align*}
The remaining probabilities are $\nu(0m1)=c(0m)-x_m$, $\nu(1m0)=c(m0)-x_m$, and $\nu(1m1)=c(1m)-c(m0)+x_m$. The fitted model corresponds to $x_m^{\mathrm{nom}}=c(0m)c(m0)/c(m)$ and, if $\eta$-positive, uniquely maximizes the entropy rate over the class.
\end{lemma}

\begin{IEEEproof}
Stationarity and compatibility with $c$ fix the row and column sums of each $2\times2$ table $\nu(\cdot\,m\,\cdot)$, leaving one free parameter $x_m$. The $\eta$-positivity constraints yield $[x_m^{\mathrm{lo}},x_m^{\mathrm{hi}}]$. The entropy rate $H(\gamma_{L+2}\mid\gamma_{1:L+1})$ is at most $H(\gamma_{L+2}\mid\gamma_{2:L+1})$, fixed by $c$, with equality iff the first and last symbols are conditionally independent given $m$, i.e., at $x_m^{\mathrm{nom}}$.
\end{IEEEproof}

The coordinates $x_m$ describe two kinds of higher-order dependencies. The coordinate $x_{0^L}=\Pr(0^{L+2})$ determines the \emph{persistence beyond the modeled memory}
\begin{equation}
\rho=\Pr(\gamma_t=0\mid S_t=0^M)
=\frac{x_{0^L}}{c(0^{L+1})},
\end{equation}
the probability that a run of more than $L$ losses continues. By \cref{lem:box}, $\rho$ ranges over $[\rho_{\mathrm{lo}},\rho_{\mathrm{hi}}]$, where $\rho_{\mathrm{lo}}=\max\{\eta,1-(1-\eta)\beta\}$, $\rho_{\mathrm{hi}}=\min\{1-\eta,1-\eta\beta\}$, and $\beta=c(10^L)/c(0^{L+1})$. The fitted model gives the nominal value $\rho_{\mathrm{nom}}=c(0^{L+1})/c(0^L)$. The other coordinates $x_m$, $m\neq0^L$, are probabilities of \emph{loss relapses}, i.e., losses that recur after at least one intervening success. For $L=1$, the only relapse coordinate is $x_1=c(01)(1-b)$, where $b=p(01)$ is the success probability after a loss followed by a success and $1-b$ is the relapse probability.

As a running example throughout the paper, we consider \mbox{$L=1$}, $M=2$, $\eta=0.05$, and $(c(00),c(01),c(10),c(11))=(0.12,0.08,0.08,0.72)$, i.e., a packet-loss rate of $0.2$ and a mean burst length of $2.5$, and, unless stated otherwise, missions of $T=60$ slots with $D=12$. Here, $x_0\in[\tfrac{11}{250},\tfrac{57}{500}]$ and $x_1\in[\tfrac1{250},\tfrac{11}{250}]$, so $\rho$ and $b$ range independently over $[\tfrac{11}{30},\tfrac{19}{20}]$ and $[0.45,0.95]$, respectively, and the fitted model has $\rho_{\mathrm{nom}}=\tfrac35$ and $b=0.9$.

\section{Dense Attempts: One Scalar Decides}
\label{sec:dense}

We first consider $B\ge T$, for which dense attempts, i.e., attempts in every slot, minimize the age pathwise and are thus optimal under every compatible channel, so the risk interval $[R^-(c),R^+(c)]$ is the range of the optimal risk over the class.

\subsection{Risk Bounds for Open-Loop Schedules}
\label{subsec:openloop}

To relate the risk to the coordinates of \cref{lem:box}, we bound it by loss-pattern probabilities for any \emph{open-loop} schedule, which attempts in a fixed set $\mathcal S\subseteq\{1,\ldots,T\}$ of slots and has risk $R(\mathcal S)$. The age reaches $D$ in slot $s+D\le T$ when a success at $s\in\mathcal S_0=\{0\}\cup\mathcal S$, or the mission start for $s=0$, is followed by losses at all attempted slots in $(s,s+D]$. We denote this event by $A_s$ and the expected number of such events by $\Lambda_{\mathcal S}=\sum_s\Pr(A_s)$, which approximates $R(\mathcal S)$ when violations are rare.

\begin{theorem}[Risk bounds for open-loop schedules]\label{thm:bound}
For any open-loop schedule $\mathcal S$,
\begin{equation}
\Lambda_{\mathcal S}(1-J_{\mathcal S}\kappa_{\mathcal S})
\le R(\mathcal S)\le\Lambda_{\mathcal S},
\end{equation}
where $\kappa_{\mathcal S}=\max_{s\in\mathcal S,\,w}\Pr(A_s\mid S_s=w)$ and $J_{\mathcal S}=\max_{s\in\mathcal S_0}|\{s'\in\mathcal S:s+D<s'\le T-D\}|$, with the maxima taken over $s$ such that $s+D\le T$.
\end{theorem}

\begin{IEEEproof}
Continue the age process beyond the first violation and let $I_t=\mathbf 1\{d_t=D\}$. Then $I_t=1$ iff $t=s+D$ for some $s\in\mathcal S_0$ and $A_s$ occurs, so $\Lambda_{\mathcal S}=\mathbb E\sum_{t=1}^T I_t$; as the age grows by one per slot, $R(\mathcal S)=\Pr(\sum_t I_t\ge1)$. For two hits at $t<t'$, the reset at $s'=t'-D$ must occur after $t$. Since $I_t$ is determined by the history through $s'-1$, the Markov property gives $\Pr(A_{s'}\mid\mathcal F_{s'-1})\le\kappa_{\mathcal S}$. Since there are at most $J_{\mathcal S}$ eligible later hits, $\sum_{t<t'}\mathbb E[I_tI_{t'}]\le J_{\mathcal S}\kappa_{\mathcal S}\Lambda_{\mathcal S}$. The union bound and the second Bonferroni inequality establish the result.
\end{IEEEproof}

\subsection{Worst-Case Persistence}
\label{subsec:worst}

With dense attempts, $\mathcal S=\{1,\ldots,T\}$, and we omit the subscript $\mathcal S$. For $M=L+1\le D$, a run of more than $L$ losses continues with probability $\rho$, so $\Pr(A_0)=\Pr(0^D)=c(0^M)\rho^k$ and $\Pr(A_s)=\Pr(10^D)=c(0^M)(1-\rho)\rho^k$ for $s=1,\ldots,n$, where $k=D-M$ and $n=T-D$. Hence
\begin{equation}
\Lambda=c(0^M)\rho^k[1+n(1-\rho)].
\label{eq:dense}
\end{equation}
Moreover, $J\kappa\le\theta_{\mathcal M}\Lambda$ with $\theta_{\mathcal M}=\max_{v:\,v_M=1}\Pr(0^M\mid v)/(c(0^M)(1-\rho))$, since $\kappa\le\rho^k\max_{v_M=1}\Pr(0^M\mid v)$ and $J=(T-2D)^+\le n$, so $0\le\Lambda-R\le\theta_{\mathcal M}\Lambda^2$ by \cref{thm:bound}. As all compatible channels share $c(0^M)$, $\rho$ thus determines the dense-attempt risk, and the extremes of $\Lambda$ over $[\rho_{\mathrm{lo}},\rho_{\mathrm{hi}}]$ locate the best and worst compatible channels, to leading order in the rare-failure regime $\theta_{\mathcal M}\Lambda\ll1$, whereas the relapse coordinates affect only higher-order terms.

\begin{theorem}[Worst-case persistence]\label{thm:worst}
For dense attempts with $k\ge1$, define $\varphi(\rho)=\rho^k[1+n(1-\rho)]$, so that $\Lambda=c(0^M)\varphi(\rho)$.
\textup{(i)} If $n\le k$, $\varphi$ is increasing, and $\Lambda$ is maximized at $\rho_{\mathrm{hi}}$.
\textup{(ii)} If $n>k$, $\varphi$ is unimodal with maximum at $\rho^\star=k(n+1)/[(k+1)n]$. Over the compatible class, $\Lambda$ is maximized at $\rho_w=\min\{\max\{\rho^\star,\rho_{\mathrm{lo}}\},\rho_{\mathrm{hi}}\}$ and minimized at an endpoint.
\textup{(iii)} The most persistent channel maximizes $\Lambda$ for every $T\ge D$ if and only if $k\ge\rho_{\mathrm{hi}}/(1-\rho_{\mathrm{hi}})$.
\textup{(iv)} For any compatible $\rho'<\rho_{\mathrm{hi}}$, $\varphi(\rho_{\mathrm{hi}})/\varphi(\rho')\ge(\rho_{\mathrm{hi}}/\rho')^k(1-\rho_{\mathrm{hi}})/(1-\rho')$. Consequently, in the rare-failure regime, $R^+/R_{\mathrm{nom}}$ and $R^+/R^-$ grow at least geometrically with $D$, with respective per-slot factors $\rho_{\mathrm{hi}}/\rho_{\mathrm{nom}}$ and $\rho_{\mathrm{hi}}/\rho_{\mathrm{lo}}$.
\end{theorem}

\begin{IEEEproof}
Claims (i)--(iii) follow from the sign of $\varphi'(\rho)=\rho^{k-1}[k(n+1)-(k+1)n\rho]$. Claim (iv) follows from $(1+nx)/(1+ny)\ge x/y$ for $0<x\le y$ and \cref{thm:bound}; see~\cite{zheng2026repo} for details.
\end{IEEEproof}

\begin{figure}[t]
\centering
\includegraphics[width=\columnwidth]{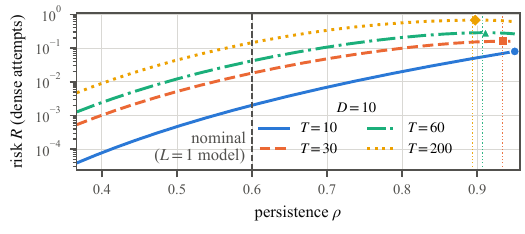}
\caption{Exact dense-attempt risk versus $\rho$, the loss probability after two losses, for several mission lengths $T$ in the running example ($D=10$, $b=0.7$; markers: exact maxima; dotted lines: $\rho^\star$ from \cref{thm:worst} for $n>k$).}
\label{fig:dense}
\vspace{-0.5cm}
\end{figure}

\Cref{thm:worst} reveals a persistence--frequency tradeoff. Increasing $\rho$ prolongs loss bursts but, for fixed $c(0^M)$, decreases their entry rate $c(0^M)(1-\rho)$. This tradeoff can place the worst case at an interior $\rho$ rather than at $\rho_{\mathrm{hi}}$, as \cref{fig:dense} shows for $D=10$. The most persistent channel is the worst there only for $T=10$, where $n\le k$, whereas for $T=60$ the exact risk peaks at $\rho=0.910$, close to $\rho^\star=0.907$, with $R^+=0.289$ against $R_{\mathrm{nom}}=0.042$ and $R^-=1.3\times10^{-3}$. The fitted model, although the maximum-entropy member of the class, is thus not generally the worst case.

\section{Budgeted Attempts: The Statistic Changes}
\label{sec:budget}

When $B<T$, the scheduler must leave some slots idle, and $\pi_L$ is in general optimal only under the fitted model. As in \cref{fig:system}, the risk depends on attempted losses, which idle slots may separate, so $\rho$ alone need not decide it.

\subsection{Periodic Attempts and Decimated Loss Runs}
\label{subsec:periodic}

Periodic attempts every $g$ slots, i.e., $\mathcal S_g=\{g,2g,\ldots\}\cap[1,T]$, are the simplest schedules with idle slots. They observe the channel only through the $g$-decimated process $\gamma_i^{(g)}=\gamma_{ig}$, with law $\Pr^{(g)}$ and run persistences $\rho_j^{(g)}=\Pr^{(g)}(0^{j+1})/\Pr^{(g)}(0^j)$, i.e., the probabilities that the next attempt fails after $j$ consecutive attempted losses.

\begin{proposition}[Periodic attempts and decimated loss runs]\label{prop:see}
For a stationary binary process $\{\gamma_t\}$ and $D\ge g$,
\begin{equation}
\Lambda_{\mathcal S_g}=\Pr\nolimits^{(g)}(0^{D_g})
\bigl[1+n_g(1-\rho_{D_g}^{(g)})\bigr],
\label{eq:decim}
\end{equation}
where $D_g=\lfloor D/g\rfloor$ and $n_g=\lfloor(T-D)/g\rfloor$. For $M=L+1$, $\rho_j^{(1)}=\rho$ for $j\ge M$, whereas $\rho_1^{(M)}=\sum_{m\in\{0,1\}^L}x_m/c(0)$.
\end{proposition}

\begin{IEEEproof}
With $A_s$ as in \cref{thm:bound}, $A_0$ requires losses at the $D_g$ attempts in $(0,D]$. Each eligible $A_s$, $s\in\mathcal S_g$, requires a success followed by $D_g$ attempted losses. Stationarity therefore gives
$\Lambda_{\mathcal S_g}=\Pr^{(g)}(0^{D_g})+n_g\Pr^{(g)}(10^{D_g})$, yielding \cref{eq:decim}. For $M=L+1$, $\Pr(0^j)=c(0^M)\rho^{j-M}$ for $j\ge M$. Moreover, $\Pr(\gamma_t=0,\gamma_{t+M}=0)=\sum_m\nu(0m0)$, which yields $\rho_1^{(M)}$.
\end{IEEEproof}

Thus, periodic attempts probe the loss-run tail of the decimated channel, as dense attempts probe that of $\gamma_t$ through $\rho$ in \cref{eq:dense}. The low-order fit fixes the first $L$ run persistences of $\gamma_t$ but, in general, not $\rho_1^{(g)}$ for $g>L$. In particular, $\rho_1^{(M)}$ combines persistence and relapse coordinates, so the relapses now enter the leading term. In the running example, varying $b$ changes the dense-attempt risk by at most about $2\%$ at any fixed $\rho$, whereas at $\rho=\tfrac{11}{30}$ lowering $b$ from $0.95$ to $0.45$ multiplies the risk of two-slot periodic attempts by $34.6$ and that of $\pi_L$ with $B=15$ by $9.05$.

The effect of spacing can also be characterized asymptotically through the matrix $Q_g(w,w')=\Pr(S_{t+g+1}=w'\mid S_{t+1}=w)$ on contexts ending in a loss, with intermediate outcomes marginalized, and its spectral radius $\sigma_g$. Here $\sigma_1=\rho$, $\sigma_g\ge\rho^g$, and $Q_g$ is entrywise positive for $g\ge M$. For fixed $g$, as $D\to\infty$, if $\log(1+T-D)=o(D)$ and either $g=1$ or $g\ge M$,
\begin{equation}
\lim_{D\to\infty}\frac{1}{D}\log R(\mathcal S_g)
=\frac{1}{g}\log\sigma_g
\label{eq:decay}
\end{equation}
by Perron--Frobenius theory and \cref{thm:bound}; see~\cite{zheng2026repo}. Per attempt, $\sigma_g$ thus takes the role of $\rho$ but also depends on the relapse coordinates, and a channel with larger $\rho$ but smaller $\sigma_g$ than another is riskier under dense attempts but safer under $g$-spaced ones for large $D$.

\subsection{Closed-Loop Scheduling}
\label{subsec:closed}

The same mechanism extends to policies that react to ACK/NACK feedback, such as $\pi_L$. For a deterministic policy $\pi$, the attempts in $(s,s+D]$ after a success at slot $s$ with observation history $h$ form a deterministic \emph{retry pattern} $\mathcal R_\pi(s,h)$ as long as they all fail. We denote by $E_{s,h,w}$ the event of a success at $s$ before any violation, with history $h$ and context $S_{s+1}=w$, where $s=0$ again represents the mission start. A partition of $\{\tau\le T\}$ by the last success before $\tau$ and the Markov property give $R(\pi)=\Lambda_\pi$, where
\begin{multline}
\Lambda_\pi=\sum_{\substack{s,h,w\\s+D\le T}}\Pr(E_{s,h,w})\\
\times\Pr\bigl(\gamma_{\mathcal R_\pi(s,h)}=0\mid S_{s+1}=w\bigr).
\label{eq:closed}
\end{multline}
In the running example with $B=15$, the retry pattern of $\pi_L$ at the mission start is $\{5,7,9,10,11,12\}$, which becomes denser as the deadline approaches, so both relapses and persistence enter its loss probability.

\subsection{Reversal of the Risk Order}
\label{subsec:reversal}

Even at moderate $D$, a budget can reverse the risk order of compatible channels. In the running example, we consider $C_1$ and $C_2$ with $(\rho,b)=(0.44,0.95)$ and $(\tfrac{11}{30},0.45)$, so $C_1$ is more persistent, whereas $C_2$ relapses more often.

\begin{proposition}[Budget reverses the risk order]\label{prop:reverse}
For $T=60$ and $D=12$, the exact risks, rounded for display, satisfy
\begin{enumerate}
\item $R(C_1)=9.10\times10^{-4}>R(C_2)=1.66\times10^{-4}$ for dense attempts ($\pi_L$, $B=60$);
\item $R(C_1)=3.46\times10^{-3}<R(C_2)=4.45\times10^{-2}$ for periodic attempts ($g=2$, $B=30$);
\item $R(C_1)=1.30\times10^{-2}<R(C_2)=6.04\times10^{-2}$ for $\pi_L$ ($B=15$).
\end{enumerate}
\end{proposition}

\begin{IEEEproof}
The risks are probabilities of absorption by slot $T$ in finite Markov chains over channel context, age, and scheduler state, computed exactly in rational arithmetic; see~\cite{zheng2026repo}.
\end{IEEEproof}

The reversal reflects the statistics each schedule observes. The run persistences of the $2$-decimated channel are lower for $C_1$ at every depth, from $\rho_1^{(2)}=0.284$ versus $0.44$ (\cref{prop:see}) to their limit $\sigma_2=0.238$ versus $0.433$, so by \cref{eq:decay} dense and two-slot periodic attempts rank $C_1$ and $C_2$ oppositely for large $D$, and for $n=T-D=48$ already for every even $D$ from $4$ to $40$. Under the fitted model, $70\%$ of $\pi_L$'s retries after a NACK are two slots apart, as in item~2).

\begin{figure}[t]
\centering
\includegraphics[width=\columnwidth]{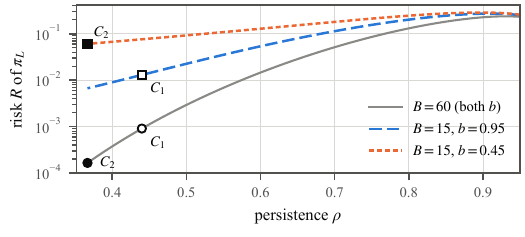}
\caption{Exact risk of $\pi_L$ versus $\rho$ for $B=60$ and $B=15$ at $b=0.45$ and $0.95$ ($T=60$, $D=12$; circles: $B=60$; squares: $B=15$).}
\label{fig:budget}
\vspace{-0.5cm}
\end{figure}

In \cref{fig:budget}, $b$ barely affects the risk with dense attempts ($B=60$) but matters with $B=15$, where $C_1$ and $C_2$ swap their order. On a $47\times21$ grid in $(\rho,b)$, reducing $B$ from $60$ to $15$ moves the grid worst case from $(0.925,0.95)$, near $\rho^\star=0.928$, to $(0.899,0.45)$, a less persistent channel that relapses more. Thus, under a budget, the loss-run tail at the attempted slots determines the risk, and no schedule-independent scalar, such as $\rho$ or a Gilbert--Elliott parameter~\cite{elliott1963}, ranks compatible channels by risk under every schedule.

\section{Numerical and Measurement-Based Evaluation}
\label{sec:case}

This section quantifies how far the low-order fit misjudges the risk over the compatible class and outside it.

\subsection{Ambiguity over the Compatible Class}
\label{subsec:class}

With dense attempts in the running example, the fitted model reports $R_{\mathrm{nom}}=1.46\times10^{-2}$ ($D=12$) and $2.07\times10^{-4}$ ($D=20$), against worst cases of $0.242$ and $0.142$ on the $47\times21$ grid of \cref{subsec:reversal}, refined in $\rho$; with $B=15$ at $D=12$, it reports $5.97\times10^{-2}$ against $0.286$. It thus underestimates the worst case by $4.8\times$ to $680\times$, and by \cref{thm:worst}(iv) the gap grows with $D$. Closing the dense-attempt gap requires an accurate estimate of $\rho$, since $\partial\log\Lambda/\partial\rho\approx14.3$ at $\rho=\rho_{\mathrm{nom}}$ and $D=12$, so an error of $0.05$ in $\rho$ changes $\Lambda$ locally by a factor of about two. For dense attempts, the closed-form certificate $\Lambda_{\max}=c(0^M)\varphi(\rho_w)$ bounds $R^+$ by \cref{thm:bound,thm:worst}; at $0.253$ and $0.143$, it exceeds the grid worst case by less than $5\%$ and $1\%$. A rectangular robust MDP~\cite{wiesemann2013} returns $0.80$--$0.98$, because it lets each context's success probability range over its projected interval and be chosen adversarially in every slot, as is the initial context, and thus discards the coupling of $p(00)$ and $p(10)$ enforced by \cref{lem:box}.

\begin{figure}[t]
\centering
\includegraphics[width=\columnwidth]{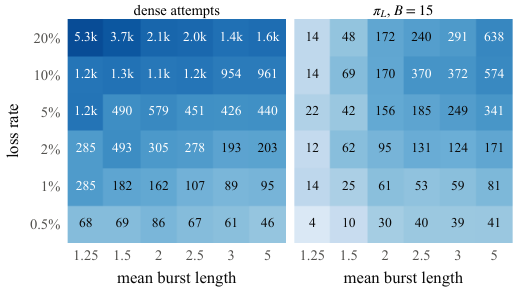}
\caption{Worst compatible-to-nominal risk ratio $R^+/R_{\mathrm{nom}}$ across Gilbert fits ($M=2$ class, $T=60$, $\eta=10^{-4}$).}
\label{fig:sweep}
\vspace{-0.5cm}
\end{figure}

In \cref{fig:sweep}, we vary the fit over loss rates from $0.5\%$ to $20\%$ and mean burst lengths from $1.25$ to $5$, each at the smallest $D$ ($6$ to $42$) for which it certifies $R_{\mathrm{nom}}\le\varepsilon=10^{-4}$ with dense attempts; the same $D$ is used for $B=15$. We use $\eta=10^{-4}$ here, since $\eta=0.05$ leaves the class empty for $27$ of these $36$ fits. With dense attempts, the worst compatible channel then has $46$ to $5.3\times10^{3}$ times the nominal risk, so $R^-\le\varepsilon<R^+$ for every fit, and with $B=15$ it has $4$ to $640$ times that of $\pi_L$. Many worst cases lie on the boundary of the class and depend on $\eta$, including the dense ones with mean burst length $5$ and loss of at least $1\%$, where $n\le k$, as in \cref{thm:worst}(i).

\subsection{Channels Outside the Class}
\label{subsec:outside}

Real channels need not have order $M=L+1$. Gilbert--Elliott channels~\cite{elliott1963}, for instance, are in general not of finite order. For each $e_B\in[0.65,1]$, the one with loss probabilities $(e_G,e_B)=(0,e_B)$ and matched transitions shares the Gilbert fit of the running example and hence $\pi_L$; $e_B=1$ gives the Gilbert channel. A loss reveals the bad state, so the persistence is $0.6$ at every burst depth, and the dense risks agree within $0.3\%$. With $B=15$, however, the risk grows from $5.97\times10^{-2}$ at $e_B=1$ to $0.127$ at $e_B=0.65$, as successes in the bad state create relapses, which lower $b$ from $0.90$ to $0.51$.

We also threshold Rayleigh fading with the Jakes spectrum~\cite{young2000} at normalized Doppler $f_D$, and composite fading at $f_D=0.03$ with log-normal shadowing of \SI{6}{dB} and a correlation time of $200$ slots, to a given loss rate. For $T=60$ and $D$ chosen per channel, \cref{tab:fading} reports the trace-driven risk and the fitted-to-trace risk ratios of \mbox{order-$L$} fits; $\pi_L$ is designed on the Gilbert fit. The error follows the run persistences of \cref{prop:see} and is two-sided. In Rayleigh fading, persistence first falls with burst depth, and the Gilbert fit is conservative. Under shadowing, it grows with depth, from $0.76$ to $0.92$ at $5\%$ loss, and for $D\ge12$ the fit underestimates the risk by a factor that grows with $D$, up to $68\times$ at $1\%$ loss and $D=16$. One more model order does not remove the error, as the second-order fit is still $19\times$ too low in that case and $1.8\times$ too low in Rayleigh fading with $f_D=0.03$ and $D=10$, where the Gilbert fit is nearly exact. Since $1-b\le0.011$, relapses are rare, and the budgeted errors follow the dense ones.

\begin{table}[t]
\centering
\caption{Trace-driven dense-attempt risk $R$ and fitted-to-trace risk ratios for fading channels and measured links 
}
\label{tab:fading}
\footnotesize
\setlength{\tabcolsep}{4pt}
\begin{tabular}{@{}lcccccc@{}}
\toprule
& & & \multicolumn{3}{c}{Dense attempts} & $\pi_L$, $B=15$ \\
\cmidrule(lr){4-6}\cmidrule(l){7-7}
Channel & Loss & $D$ & $R$ & $L{=}1$ & $L{=}2$ & $L{=}1$ \\
\midrule
Rayleigh, $f_D{=}0.01$ & $5\%$ & 20 & $0.018^\dagger$ & 1.7 & 1.5 & 1.7 \\
Rayleigh, $f_D{=}0.03$ & $5\%$ & 10 & 0.025 & 1.1 & 0.54 & 1.4 \\
Shadowed & $5\%$ & 18 & 0.016 & 0.30 & 0.32 & 0.40 \\
Shadowed & $1\%$ & 12 & 0.0025 & 0.086 & 0.21 & 0.19 \\
Shadowed & $1\%$ & 16 & 0.0010 & 0.015 & 0.053 & 0.034 \\
FlockLab & $12\%$ & 15 & 0.050 & 0.17 & 0.39 & 0.26 \\
FlockLab & $19\%$ & 18 & 0.080 & 0.10 & 0.26 & 0.18 \\
\bottomrule
\multicolumn{7}{@{}l}{\scriptsize $^\dagger$Below $R^-$ of the $M=2$ class ($\eta=10^{-4}$); all other risks lie in $[R^-,R^+]$.}
\end{tabular}
\vspace{-0.5cm}
\end{table}

We further use IEEE 802.15.4 traces of the FlockLab testbed~\cite{jacob2019flocklab}, namely $100$-strobe tests on the $171$ directed TelosB links with $1$--$25\%$ loss and at least $500$ tests, with one slot per strobe and missions sliding within tests. The records are too short for $10^{-4}$, so $D$ is the smallest survival time at which the Gilbert fit certifies $R_{\mathrm{nom}}\le10^{-2}$ with dense attempts. The measured dense risk then exceeds the certified $10^{-2}$ on $55\%$ of the links and the nominal risk on $64\%$, by up to $12\times$, and by more than two standard errors over tests on $42\%$ and $47\%$, respectively. It is significantly lower on $20\%$, all of which have mean burst length below $1.9$, and the second-order fit is too low on $85\%$, significantly on $49\%$. With $B=15$, the nominal risk of $\pi_L$ is significantly too low on $33\%$ of the links, by up to $5.4\times$, and significantly too high on $36\%$. The dense error grows with loss rate and burstiness, with rank correlations of $0.93$ and $0.92$, and the $M=2$ class brackets the measured dense risk on the burstiest link of each of four loss bands, including the two links in \cref{tab:fading}. Variation between the pooled tests may also contribute to these errors.
    
\raggedbottom
\section{Conclusion} \label{sec:conclusion}
This paper examined how far low-order packet-loss statistics determine survival-time risk under acknowledgment-based scheduling with a transmission budget, and bounded this risk for every channel in the compatible class of higher-order Markov channels. With dense attempts, the persistence beyond the modeled memory decides the risk to first order; as higher persistence makes bursts longer but rarer, the most persistent channel need not be the worst. With a limited budget, spaced retries bring relapse probabilities into the risk and can reverse the risk order of compatible channels. Compatible channels can be orders of magnitude riskier than the fit predicts, and on fading channels and measured links the fit can underestimate the risk even with one more model order. Risk should therefore be assessed from the loss-run tail that the schedule observes, not from the fit. The structural results assume channels one order above the fit; for channels of unknown order, estimating this tail from limited traces is left for future work.
 
\bibliographystyle{IEEEtran}
\bibliography{IEEEabrv,refs}

\end{document}